# Nature of the finite temperature transition in QCD with strange quark *

Y. Iwasaki[a], K. Kanaya[a], S. Kaya[a], S. Sakai[b] and T. Yoshié[a]

[a]Institute of Physics, University of Tsukuba, Ibaraki 305, Japan

[b]Faculty of Education, Yamagata University, Yamagata 990, Japan

The finite temperature transition in QCD is studied using Wilson quarks for the cases of $N_F = 2$, 3 and 2+1. For $N_F = 2$ the transition is smooth in the chiral limit on both $N_t = 4$ and 6 lattices. For $N_F = 3$, clear two state signals are observed for $\beta \leq 4.7$ on $8^2 \times 10 \times 4$ and $12^3 \times 4$ lattices, which implies the transition is first order for $m_q \lesssim 140$ MeV. For $N_F = 2 + 1$ we study two cases of $m_s \simeq 150$ and 400 MeV with $m_u = m_d \simeq 0$. In contrast to a previous result with staggered quarks, two state signals are clearly observed for both cases, suggesting a first order QCD phase transition in the real world.

## 1. INTRODUCTION

The study of the finite temperature QCD transition provides important information for heavy ion collisions and the evolution of early Universe: Among others the order of the transition is the most decisive information. Previous investigations show that the order of the chiral transition depends sensitively on the number of flavors $N_F$: The transition is first order for $N_F \geq 3$, while a continuous transition is suggested for $N_F = 2$. Because the s-quark mass is of the same order of magnitude as the transition temperature, it is necessary to include s-quark properly to make a prediction for the real world.

In this report we study the finite temperature transition with Wilson quarks. We first present the results of the simulations for $N_F = 2$ and 3 which are the extensions of our previous studies[1,2]. We then discuss a more realistic case of $N_F = 2 + 1$: two degenerate light u and d-quarks and a heavier s-quark. Comparison with previous results with staggered quarks is also made. We mainly perform simulations on $8^2 \times 10 \times N_t$ and $12^3 \times N_t$ lattices with $N_t$ (lattice size in the temporal direction) = 4. The lattice size 10 and 12 in one direction is doubled for hadron calculations. For $N_F = 2$ we also study the case of $N_t = 6$. The method of simulations is similar to that described in [1].

## 2. $N_F = 2$

For the case of $N_F = 4$ we showed in [2] that the finite temperature transition line, $K_T$-line, crosses the chiral limit line, $K_C$-line, at $\beta = \beta_{CT} \simeq 3.9 - 4.0$ and that the transition is smooth in the chiral limit.

In this paper, we extend the study to the case of $N_t = 6$. Our result is shown in Fig. 1. Similar to the case of $N_t = 4$, we find that, when we decrease $\beta$ toward $\beta_{CT} \simeq 4.0 - 4.2$, $m_\pi^2$ decreases rapidly and is consistent with zero at $\beta_{CT}$. This suggests a continuous transition.

## 3. $N_F = 3$

Appling the same method as in the case of $N_F = 2$ to the case of $N_F = 3$ at $N_t = 4$, we observed a two-state signal at $\beta_{CT} \simeq 3.0$[2]: When we start from a hot configuration, it remains stable in the deconfining phase. However, when we start from a mixed initial configuration, the plaquette, the Polyakov loop and $m_\pi^2$ decrease rapidly with MC-time, showing that the system is developing into a confining state. This result together with the first order signal observed for $N_F = 6$ [1] is consistent with the prediction based on universality [3].

We here extend the study to finite $m_q$'s by increasing $\beta$ on the $K_T$-line. Our phase diagram

*Talk presented by K. Kanaya at *Lattice 94*, Bielefeld, Germany. To appear in the proceedings.

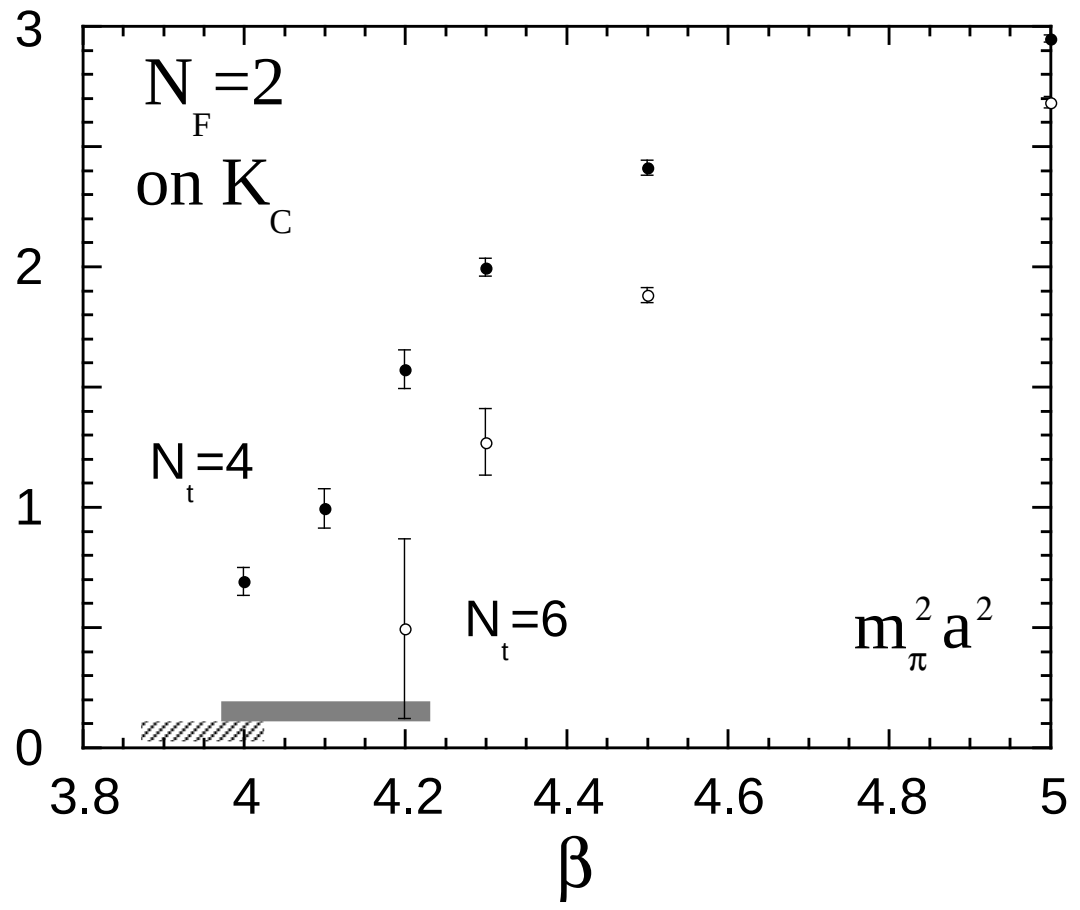

Figure 1. $(m_\pi a)^2$ on the $K_C$-line for $N_t = 4$ and 6 with spatial size $8^2 \times 10$. The locations of the crossing point $\beta_{CT}(N_t)$ are shown by shaded bars.

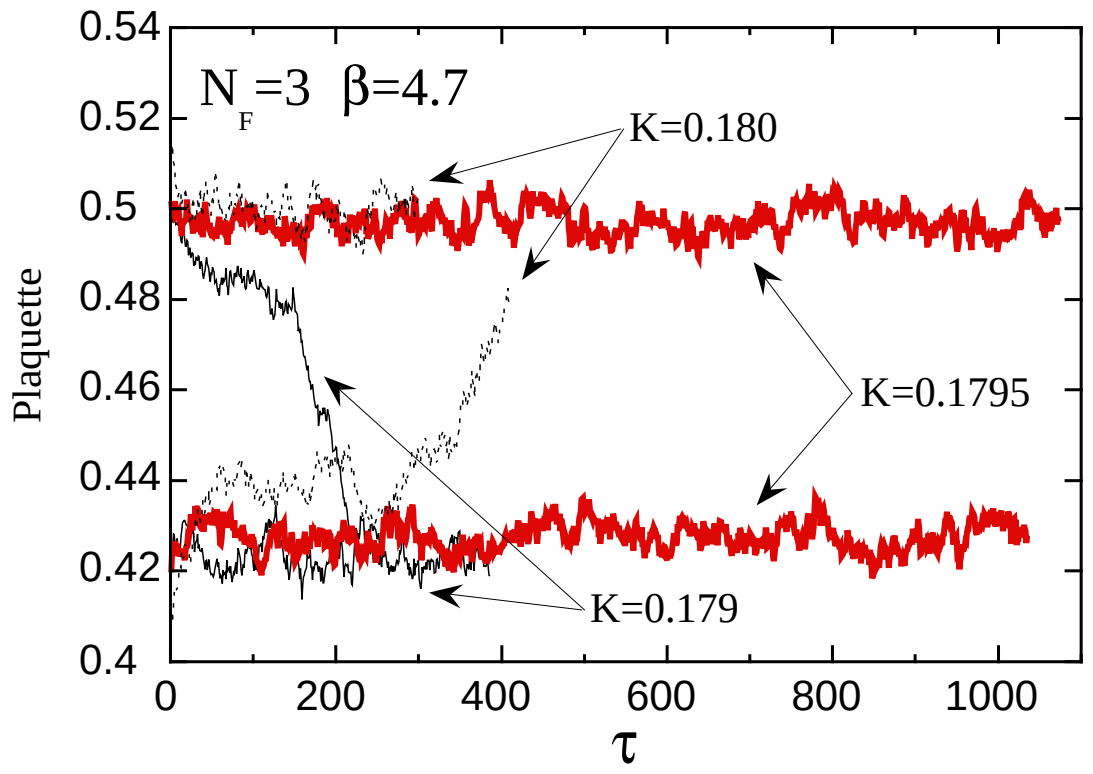

Figure 2. Time histories of the plaquette for $N_F = 3$ at $\beta = 4.7$ on a $12^3 \times 4$ lattice.

thus obtained is given in [4]. We find clear two-state signals for $\beta = 4.0$, 4.5 and 4.7 on both $8^2 \times 10 \times 4$ and $12^3 \times 4$ lattices (Fig. 2). However, when we further increase up to $\beta \geq 5.0$, we find no clear signs of metastability. The critical value of the quark mass $m_q^{\rm crit}$, up to which we get a clear first order signal on these lattices, is bounded from below by $m_q$ for $\beta = 4.7$.[1] We find $m_q^{\rm crit} a \geq 0.175(2)$ and $(m_\pi/m_\rho)^{\rm crit} \geq 0.873(6)$.

In order to convert the critical quark mass into physical units, we use the results of $m_\rho a$ and $m_q a$ for these $\beta$'s. When we plot the masses in the confining phase as a function of $1/K - 1/K_C$, they are almost independent of $\beta$, $N_F$ and $N_t$ for $\beta \lesssim 4.7$. Fig. 3 shows our results for $N_F = 2$ and 3. Identifying $m_\rho(K_C)$ with 770 MeV we get $a \simeq 0.8$ GeV$^{-1}$ for $\beta \lesssim 4.7$. It should be noted that the vector meson mass for $m_q = m_s \simeq 150$ MeV is consistent with $m_\phi = 1020$ MeV, indicating that our choice of the definition of the quark mass is close to that of the current quark mass.

From this value of $a$ we get $m_q^{\rm crit} \gtrsim 140$ MeV. We note that our $m_q^{\rm crit}$ is much larger than that obtained previously for staggered quarks at $N_t = 4$ with spatial $8^3$ – $16^3$ lattices [8,9]: two state signals are observed for $m_q a = 0.025$, while no clear metastabilities are found for $m_q a = 0.075$. Using a result for meson masses at corresponding $\beta$ [10] (the values for $N_F = 4$, because they are only available) we obtain $m_q^{\rm crit} \simeq 12$ – 38 MeV and $(m_\pi/m_\rho)^{\rm crit} \simeq 0.42$ – 0.58 for staggered quarks.

## 4. $N_F = 2+1$

Let us now turn our attention to the case of $N_F = 2+1$: $m_u = m_d < m_s$. In order to see what happens with realistic quark masses, we study the cases $m_s \simeq 150$ MeV and 400 MeV. Corresponding values of $K_s$ are obtained from Fig. 3. Keeping $m_u = m_d \simeq 0$ we decrease $\beta$ until we hit $K_T$. We find two state signals on an $8^2 \times 10 \times 4$ lattice for both $m_s = 150$ and 400 MeV and also on a $12^3 \times 4$ lattice for $m_s = 400$ MeV (Fig. 4).

With staggered quarks, Columbia group [9] reported that no transition occurs at $m_u a = m_d a = 0.025$, $m_s a = 0.1$ ($m_u = m_d \simeq 12$ MeV, $m_s \simeq 50$ MeV using the value of $a$ from [10]). Note in this connection that if $m_s < m_q^{\rm crit}$ (for degenerate $N_F = 3$), we should certainly expect a first order transition. Therefore when we recall $m_q^{\rm crit} \gtrsim 140$ MeV for Wilson quarks and $m_q^{\rm crit} \simeq 12$ – 38 MeV for staggered quarks, the results for $N_F = 2 + 1$ with Wilson and staggered quarks are both consistent with the results for degenerate $N_F = 3$ cases, respectively.

---
[1] For $m_q$ we use that in the confining phase with the definition of $m_q$ by an axial-vector Ward identity [5,6]. For $\beta \lesssim 5.3$, we experience that $m_q$ shows strange behavior in the deconfining phase [1]. This can be attributed to an effect of $O(a)$ chiral violation of Wilson fermions. In the confining phase, on the other hand, no strange behavior is found. A study with an RG improved action shows that the masses in the confining phase suffer almost no $O(a)$ effects [7].

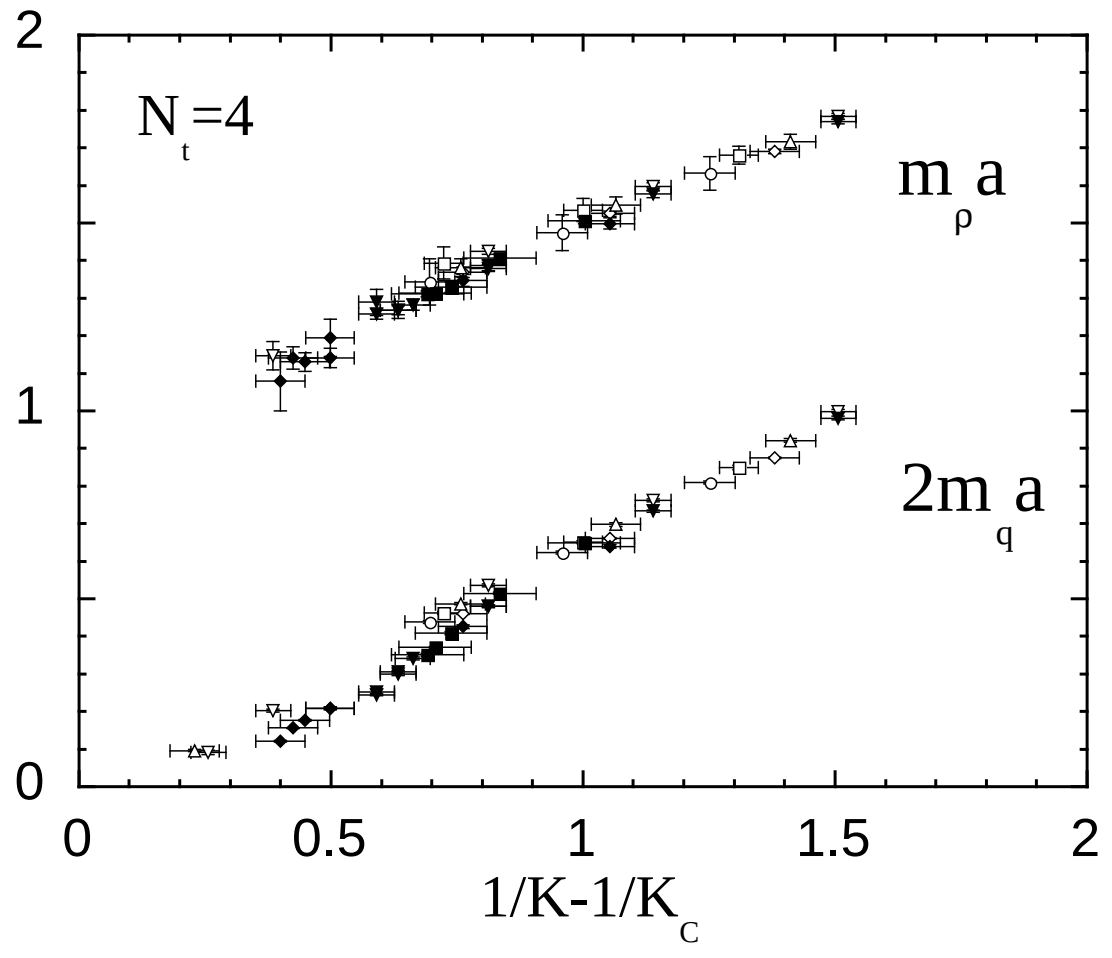

Figure 3. $m_\rho a$ and $m_q a$ in the confining phase. Open symbols are for $N_F = 2$, $\beta = 3.0$, 3.5, 4.0, 4.3, and 4.5 on an $8^2 \times 10 \times 4$ lattice. Filled symbols are for $N_F = 3$, $\beta = 4.0$, 4.5 and 4.7 on $8^2 \times 10 \times 4$ and $12^3 \times 4$. $K_C(\beta)$ is determined by $m_\pi^2$ and $m_q$ for $N_F = 2$, with errors taking into account the difference between two definitions.

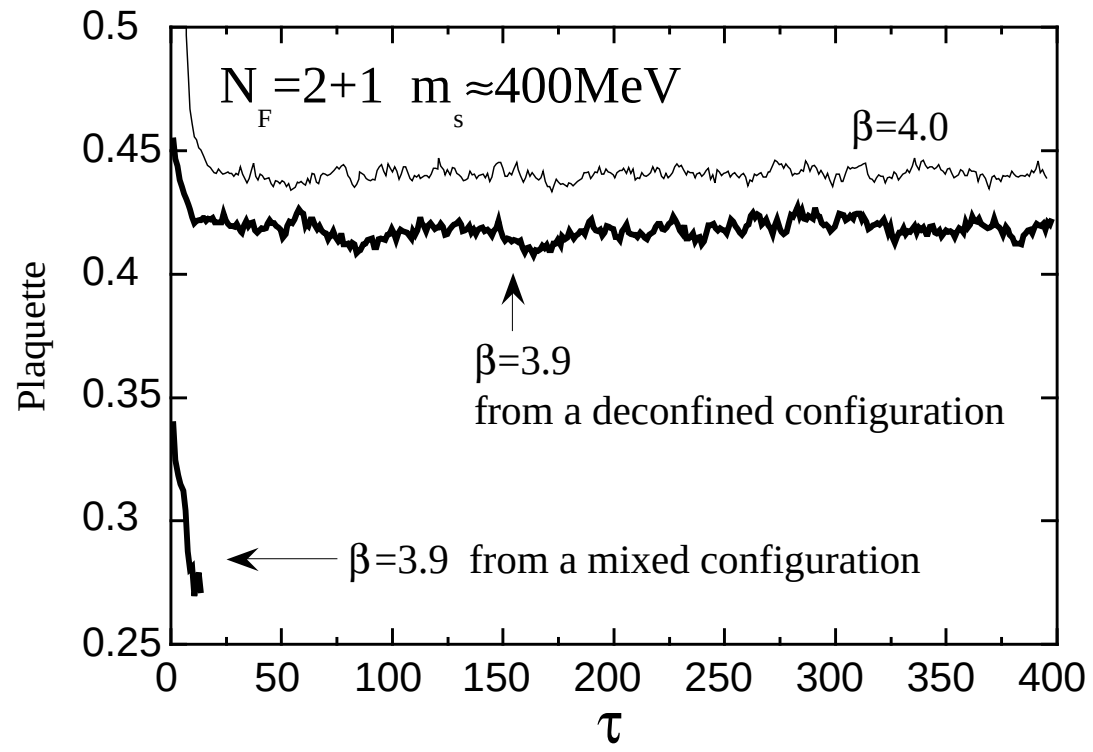

Figure 4. Time histories of the plaquette for $N_F = 2 + 1$ with $m_s \simeq 400$ MeV on a $12^3 \times 4$ lattice.

## 5. CONCLUSION

We have studied the nature of the finite temperature transition with Wilson quarks for $N_F =$ 2, 3 and 2+1. For $N_F = 2$ the chiral transition is smooth on an $N_t = 6$ lattice in accord with our previous result at $N_t = 4$. For $N_F = 3$ at $N_t = 4$, clear two state signals are observed on the $K_T$-line for $m_q \lesssim 140$ MeV. For $N_F = 2 + 1$ we have studied the cases $m_s \simeq 150$ and 400 MeV with $m_u = m_d \simeq 0$, and we have found two state signals for both cases. This suggests a first order finite temperature transition in the real world.

The simulations are performed with HITAC S820/80 at KEK and with VPP-500 at the University of Tsukuba. We thank members of KEK for their support. We are grateful to C. DeTar and K. Rajagopal for useful discussions. This work is in part supported by the Grant-in-Aid of Ministry of Education, Science and Culture (No.06NP0601).